# Instant determination of the potential biomarker heparan sulfate in human plasma by a mix-and-read fluorescence assay

Ulrich Warttinger[1], Roland Krämer[1]

Correspondence to:
Roland Krämer, phone 0049 6221 548438, fax 0049 6221 548599
E-mail: kraemer@aci.uni-heidelberg.de

1 Heidelberg University, Inorganic Chemistry Institute, Im Neuenheimer Feld 270, 69120 Heidelberg, Germany.

## Abstract

Heparan sulfate (HS) is a linear, polydisperse sulfated polysaccharide belonging to the glycosaminoglycan family. HS proteoglycans are ubiquitously found at the cell surface and extracellular matrix in animal species. HS is involved in the interaction with a wide variety of proteins and the regulation of many biological activities. In certain pathologic conditions, expression and shedding of HS proteoglycans is overregulated, or enzymatic degradation of HS in lysosomes is deficient, both leading to excess circulating free HS chains in blood plasma. HS has therefore been suggested as a biomarker for various severe disease states. The structural heterogeneity makes the quantification of heparan sulfate in complex matrices such as human plasma challenging. HS plasma levels are usually quantified by either disaccharide analysis or enzyme linked immunosorbent assay (ELISA). Both methods require time-consuming multistep-protocols. We describe here the instant detection of heparan sulfate in spiked plasma samples by the Heparin Red Kit, a commercial mix-and–read fluorescence microplate assay. The method enables HS quantification in the low µg/mL range without sample pretreatment. Heparin Red appears to be sufficiently sensitive for the detection of highly elevated HS levels as reported for mucopolysaccharidosis, graft versus host disease after transplantation, dengue infection or septic shock. This study is a significant step toward the development of a convenient and fast method for the quantification of HS in human plasma, with the potential to simplify the detection and advance the acceptance of HS as a biomarker.

## Keywords

## Introduction

*HS structure and biological role*

Heparan sulfate (HS) is a linear, polydisperse sulfated polysaccharide belonging to the glycosaminoglycan family. It is built of a major repeating disaccharide unit consisting of glucoronic acid and glucosamine (scheme 1) [1, 2]. The glucosamine may be 6-O-sulfated, either N-sulfated or N-acetylated, and sporadically 3-O-sulfated, while the glucuronic acid may be 2-O-sulfated. Typically, the average degree of sulfation is around one per disaccharide. HS proteoglycans are ubiquitously found at the cell surface and extracellular matrix in animal species. HS interacts with a large number of proteins and is involved in the regulation of many biological activities including cell proliferation, inflammation, angiogenesis, viral infectivity and development [3, 4].

*Pathologic elevation of HS plasma levels and HS as potential biomarker*

In certain disease states, expression and shedding of HS proteoglycans is overregulated or enzymatic degradation of HS in lysosomes is deficient, both leading to excess circulating free HS chains in blood plasma that may be partially excreted into urine. Thus, elevated HS levels in plasma, serum or urine have been suggested as a biomarker for a variety of disease states, including various types of mucopolysaccharidosis, graft versus host disease (GVHD), dengue infection and septic shock. While normal plasma displays a low HS level in the range 0,1-0,3 µg/mL (as determined by disaccharide analysis, table 1), pathologic HS levels are often multifold enhanced (table 2).

Biomarkers as measurable indicators of normal biological processes, pathogenic processes or response to a therapeutic intervention, are becoming increasingly important in diagnosis and drug development. Among the large number of potential biomarkers, only few have been approved to date by the US FDA. The availability of reliable quantitation techniques is crucial for the verification and validation of putative biomarkers.

The potential of HS as a biomarker is exemplified by graft versus host disease (GVHD), a life-threatening complication related to transplantation. After hematopoietic stem cell transplantation, serum HS is highly elevated *before* the onset of clinical symptoms of GVHD, and the HS serum level correlates to the severity of the disease [5]. The early diagnosis and treatment (by serum protease inhibitors) of GVHD has been shown to improve survival. Therefore, elevated serum or plasma HS levels may serve as an early biomarker of acute cellular rejection, and HS level is a relevant parameter for therapy monitoring.

| Quantification Method | HS plasma level (µg/mL) | charge per monosaccharide | Reference |
|---|---|---|---|
| Carbazole | average 0,2<br>range 0,1-0,3 (n=6) | -1.1 | [6] |
| DS analysis | n.d. | -0,83 [a]<br>(n=3, male AB serum)<br>-1,0<br>(n=3, serum from platelet poor plasma) | [7] |
| DS analysis | range 0,1-0,3 (n=35) | n.d. | [8] |
| DS analysis | range 0,1-0,2 (n=4) | -0.55 [a] | [9] |
| DS analysis | range 0,2-0,3 (n=26)<br>(serum, pre-menopausal)<br>range 0,1-0,2 (n=25)<br>(serum, post-menopausal) | -0,97 [a]<br>(pre-menopausal)<br>-0,91<br>(post-menopausal) | [10] |
| DS analysis | average ca. 0,2 (n=16) | -0.66 [a] | [11] |
| ELISA | average 4,4<br>range 0,3-9,8 (n=450) | n.d. | [12] |
| ELISA | average 5,9<br>range 4,8-7,9 (n=10) | n.d. | [13] |
| ELISA | average ca. 7<br>range ca. 5-40 (n=24) | n.d. | [14] |

Table 1. HS concentration and charge in plasma or serum of healthy donors (n: number of donors), measured by the carbazole method, disaccharide (DS) analysis, or enzyme linked immunosorbent assay (ELISA). All ELISA quantifications used the same commercial kit from the manufacturer Seikagaku, Nr. 280564. n.d. means "not detected". [a] Derived from the relative abundance of unsulfated, mono-, di- and trisulfated disaccharides as given in the reference.

*Methods for HS quantification in plasma or serum*

The structural heterogeneity makes the quantification of heparan sulfate in complex matrices such as human plasma challenging. HS level in human plasma or serum is usually quantified by either disaccharide analysis (DS analysis) or enzyme linked immunosorbent assay (ELISA), see table 1 and table 2 for examples. DS analysis provides not only the HS quantity but also valuable information on HS structure. The method is labour-intensive, time-consuming and requires expensive instrumentation. The tedious multistep-protocols typically include extensive plasma sample pretreatment, followed by isolation of HS from plasma by chromatographic methods, enzymatic digestion of HS oligomers into disaccharides, eventually a modification step such as fluorophore labelling, and finally the identification and quantification of disaccharides by LC-MS methods or electrophoresis. Earlier protocols [6] have described colorimetric quantification of HS by uronic acid- carbazole reaction, but also require a complex procedure for the extraction and purification of HS from plasma.

| Disease state | Quantification method | HS plasma level (µg/mL) | charge per monosaccharide | Reference |
|---|---|---|---|---|
| MPS I | DS analysis | range 0,35-0,65 (n=6) | n.d. | [8] |
| Indirect lung injury | DS analysis | range 0,4- ca. 10 (n=5) | -0.73 [a] | [9] |
| Rheumatoid arthritis | DS analysis | range 0,2-0,4 (n=25) (serum, pre-menopausal) range 0,3-0,4 (n=25) (serum, post-menopausal) | -1,07 [a] (pre-menopausal) -1,13 (post-menopausal) | [10] |
| MPS II | ELISA | average 80 range 4-203 (n=26) | n.d. | [12] |
| Ischemia/early reperfusion [b] | ELISA | average 112 (n=5) | n.d. | [13] |
| GVHD grade III-IV | ELISA | average 24 (n=11) (serum) | n.d. | [5] |
| Dengue [c] | ELISA | average 146 (n=75) range 87-209 | n.d. | [15] |
| Septic shock | ELISA | average ca. 30 (n=24) range ca. 6-200 | n.d. | [14] |

Table 2. HS concentration and charge in plasma or serum of patients (n: number of patients), measured by disaccharide (DS) analysis, or enzyme linked immunosorbent assay (ELISA). All ELISA quantifications used the same commercial kit from the manufacturer Seikagaku, Nr. 280564. n.d. means "not detected". MPS I / II means "mucopolysaccharidosis type I / II". GVHD means "graft versus host disease". [a] Derived from the relative abundance of unsulfated, mono-, di- and trisulfated disaccharides as given in the reference. [b] Patients undergoing cardiopulmonary bypass. [c] day 5-6 of illness.

The ELISA method is based on two heparan sulfate-specific monoclonal antibodies that target N-acetylated and N-sulfated sites of HS. It is easier to perform than disaccharide analysis and requires "only" a microplate reader as spectroscopic instrumentation. The commonly used ELISA method still requires sample pre-teatment and long incubation times, so that overall assay time HS detection in serum sums up to about 40 hours according to the protocol of the provider [16]. The ELISA assay commonly detects a much higher (by a factor 20-40) HS level in human plasma than the disaccharide assays (table 1). All detections listed in table 1 used a HS ELISA kit from the same manufacturer. A reasonable explanation is therefore that the HS calibrant for construction of the standard curve has a lower number of epitopes than human plasma HS [14].

In one study, the HS content of 18 urine samples (4 samples of normal donors and 14 samples of MPS and other patients) was analyzed by both methods: Disaccharide analysis and ELISA [12]. The correlation between the two methods was poor, the ELISA/HPLC ratio of detected HS ranged between 0.26 and 15.8, i.e. in some samples, ELISA detects a 4 times lower HS quantity than disaccharide analysis, while in other samples, ELISA detects 16 times more HS than disaccharide analysis. The discrepancy may be related to the structural diversity of HS molecules.

Assuming that normal HS plasma levels are about 0,2 µg/mL as detected by disaccharide analysis (table 1), a tentative "correction" of the ELISA-determined concentrations, based on the relative increase of HS over controls or normal levels, yields HS plasma levels in the range 1-5 µg/mL for mucopolysaccharidosis, dengue infection, graft versus host disease and septic shock (table 2).

A convenient and fast method for the quantification of heparan sulfate in human plasma or serum would simplify the detection of HS and advance its use as a biomarker. We describe here as an alternative method for HS detection in human plasma the application of the fluorescent probe "Heparin Red". The latter was initially described as an experimental probe for heparin detection [17] and has meanwhile been developed further into commercially available assays [18]. Heparin Red is a polyamine modified, red-emissive perylene diimide fluorophore (scheme 1, right). It forms a supramolecular aggregate with polysulfated polysaccharides, resulting in contact quenching of fluorescence (scheme 2). The strong binding of the polycationic probe to polyanionic heparin appears to be controlled by both electrostatic and aromatic pi-stacking interactions [19]. In this contribution, we describe for the first time application of Heparin Red to the determination of heparan sulfate in human plasma. This is to our knowledge the first example for the direct detection of HS plasma levels in the low µg/mL range by a mix-and-read optical assay.

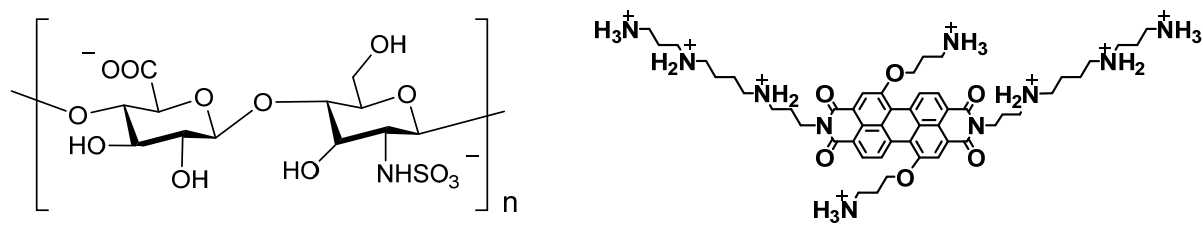

**Scheme 1**. Left: Major glucuronic acid - glucosamine repeating disaccharide unit of heparan sulfate, mono-sulfated form. The sulfation pattern is highly variable, possible sites of sulfation are glucuronic acid O2 and glucosamine O6 and N (either sulfated or acetylated). The disaccharide moiety may be unsulfated, monosulfated, disulfated or (rarely) trisulfated. Right: Structure of the polycationic fluorescent probe Heparin Red.

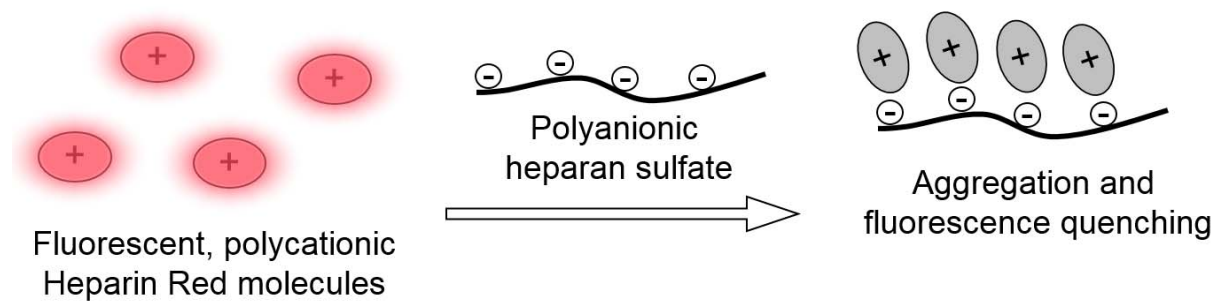

**Scheme 2.** Schematic representation of fluorescence quenching of the molecular probe Heparin Red in the presence of heparan sulfate due to formation of non-fluorescent aggregates.

## Materials and Methods

**Instrumentation**

*Fluorescence measurements*

Fluorescence was measured with a microplate reader Biotek Synergy Mx (Biotek Instruments, Winooski, VT, USA), excitation at 570 nm, emission recorded at 605 nm, spectral band width 13,5-17 nm, gain 90-110 , read height of 8 mm.

*Microplates*

For fluorescence measurements 96 well microplates, polystyrene, Item No 655076, were purchased from Greiner Bio-One GmbH, Frickenhausen.

*Pipettes*

Transferpette® 0,5-10µl, Transferpette®-8 20-200µl and Transferpette®-12 20-200µl, purchased from Brand GmbH, Wertheim. Rainin Pipettes 100-1000µl, 20-200µl, and 2-20µl purchased from Mettler Toledo, OH, USA.

**Reagents**

*Heparin Red® Kit and Heparin Red® Ultra*

The Heparin Red® Kit was a gift from Redprobes UG, Münster, Germany [18]. Kit components: Heparin Red solution, Product No HR001, Lot 01-003, and Enhancer Solution, Product No ES001, Lot 004. Heparin Red® Ultra was a gift from Redprobes UG, Münster, Germany [18]. Product No HRU001, Lot 003.

*Heparan sulfate*

Heparan sulfate from porcine mucosa, highly purified fractions I and III (Cat No GAG-HS I and GAG-HS III) were purchased from Iduron Ltd, Cheshire, UK.

*Plasma*

Five single-donor citrated plasmas of healthy individuals were provided by the Blood Bank of the Institute for Clinical Transfusion Medicine and Cell Therapy Heidelberg (IKTZ). The single donor plasmas were applied as matrix for the detections shown in figures 4 and 5. The heparan sulfate spiked plasmas were stored at -20°C. Pooled normal plasma, applied as matrix for the detections shown in figure 3, was prepared by mixing equal volumes of the five single donor plasmas.

*Other*

Aqueous solutions were prepared with HPLC grade water purchased from VWR, product No 23595.328..

**Assays**

*Heparin Red® Kit*

For determination of heparan sulfate concentrations in plasma or aqueous samples, the protocol of the provider for a 96-well microplate assay was followed. Mixtures of Heparin Red solution and Enhancer solution (volume ratio 1:90) were freshly prepared. 20 µL of the heparan sulfate spiked plasma sample (or heparan sulfate containing aqueous samples) was pipetted into a microplate well, followed by 80 µL of the Heparin Red – Enhancer mixture. For sample numbers > 10, a 12-channel pipette was used for addititon of the Heparin Red – Enhancer solution. The microplate was introduced in the fluorescence reader and mixing was performed using the plate shaking function of the microplate reader (setting "high", 3 minutes). Immediately after mixing, fluorescence was recorded within 1 minute.

*Heparin Red® Ultra*

For determination of low heparan sulfate concentrations in aqueous solution, the protocol of the provider for a 96-well microplate was modified: 90 µL of the heparan sulfate containing sample was pipetted into a microplate well. Then, 90 µL of Heparin Red Ultra solution was added to the samples as simultaneously as possible. For sample numbers > 10, a 12-channel pipette was used for addititon of Heparin Red Ultra solution. The microplate was immediately introduced in the fluorescence reader and mixing was performed using the plate shaking function of the microplate reader (setting "high", 1 minute). Immediately after mixing, fluorescence was recorded within 1 minute. Detections were performed as duplicates.

*Spiked plasma samples*

Plasma samples containing defined mass concentrations of heparin sulfate were prepared by adding aqueous solutions (10 vol%) of the corresponding heparan sulfate to plasma and vortexing, to achieve a concentration of 100 µg/mL. Heparan sulfate concentrations required for the detections were adjusted by further dilution of this 100 µg/mL stock solution with the same plasma and vortexing. The spiked plasma samples were stored at -20°C for at least one day and up to several months, thawed at room temperature and vortexed before use.

# Results and discussion

*Detection of heparan sulfate in aqueous samples*

All detections have been performed with spiked samples using commercially available, highly purified heparan sulfate fractions (HS I and HS III) from porcine mucosa. Selected properties of HS I and HS III are given in table 3. The fractions differ significantly in molecular weight and sulfation degree [20]. The negative charge density per monosaccharide of the HS fractions is significantly lower in comparison to heparin (-1,7).

| HS fraction | Average mol. weight | Average sulfation (per disaccharide) | Average charge density (per monosachharide) |
|---|---|---|---|
| HS I | 40 kD | 0,75 | -0,88 |
| HS III | 9 kD | 1,75 | -1,38 |

Table 3. Selected properties of heparan sulfate fractions HS I and HS III [20].

The Heparin Red Kit was first applied to the detection of heparan sulfate at different concentrations in aqueous samples, using the protocol of the provider for plasma samples. Figure 1 shows the decrease of fluorescence signal with increasing heparan sulfate concentration. The lower the charge density of the HS, the more HS is needed for fluorescence quenching. The ratio of HS III/ HS I monosaccharide charge density (table 1) is 1,57 what matches well the 1,5-fold higher amount of HS I required to achieve the same fluorescence quenching as with HS III (figure 1). This confirms the suggested formation of charge neutral, non-fluorescent aggregates between Heparin Red and sulfated polysaccharides in non-competitive matrices [19].

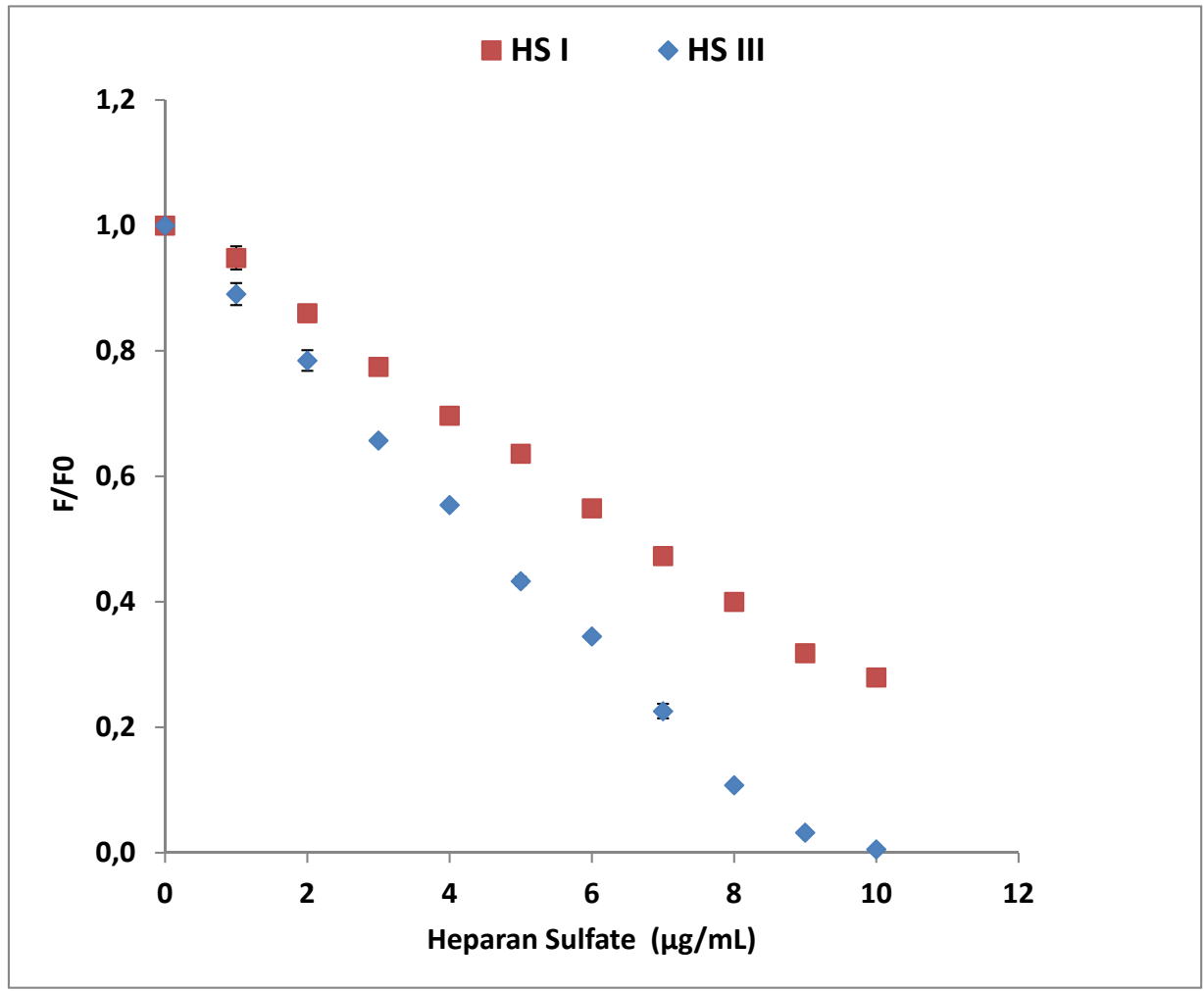

**Figure 1.** Normalised response curve of the Heparin Red Kit using aqueous heparan sulfate samples HS I and HS III in the concentration range 0-10 µg/mL (following protocol of the provider). Manually performed microplate assay. Excitation at 570 nm, fluorescence emission at 605 nm. Averages of duplicate determinations; CV's (averaged over all concentrations) for the linear range 0 to ≈9 µg/mL of HS I and HS III 2,7 %.

A higher sensitivity can be achieved when a different commercial assay, Heparin Red Ultra, is applied. This assay is preferable for matrices other than plasma or serum [18, 21] and can be more readily adapted for higher sensitivity. Using a modified protocol, the detection limit for heparan sulfate is about 0,1 µg/mL (figure 2). These results suggest that sensitivities in the sub- µg/mL range can be achieved in less complex matrices using the Heparin Red Ultra assay.

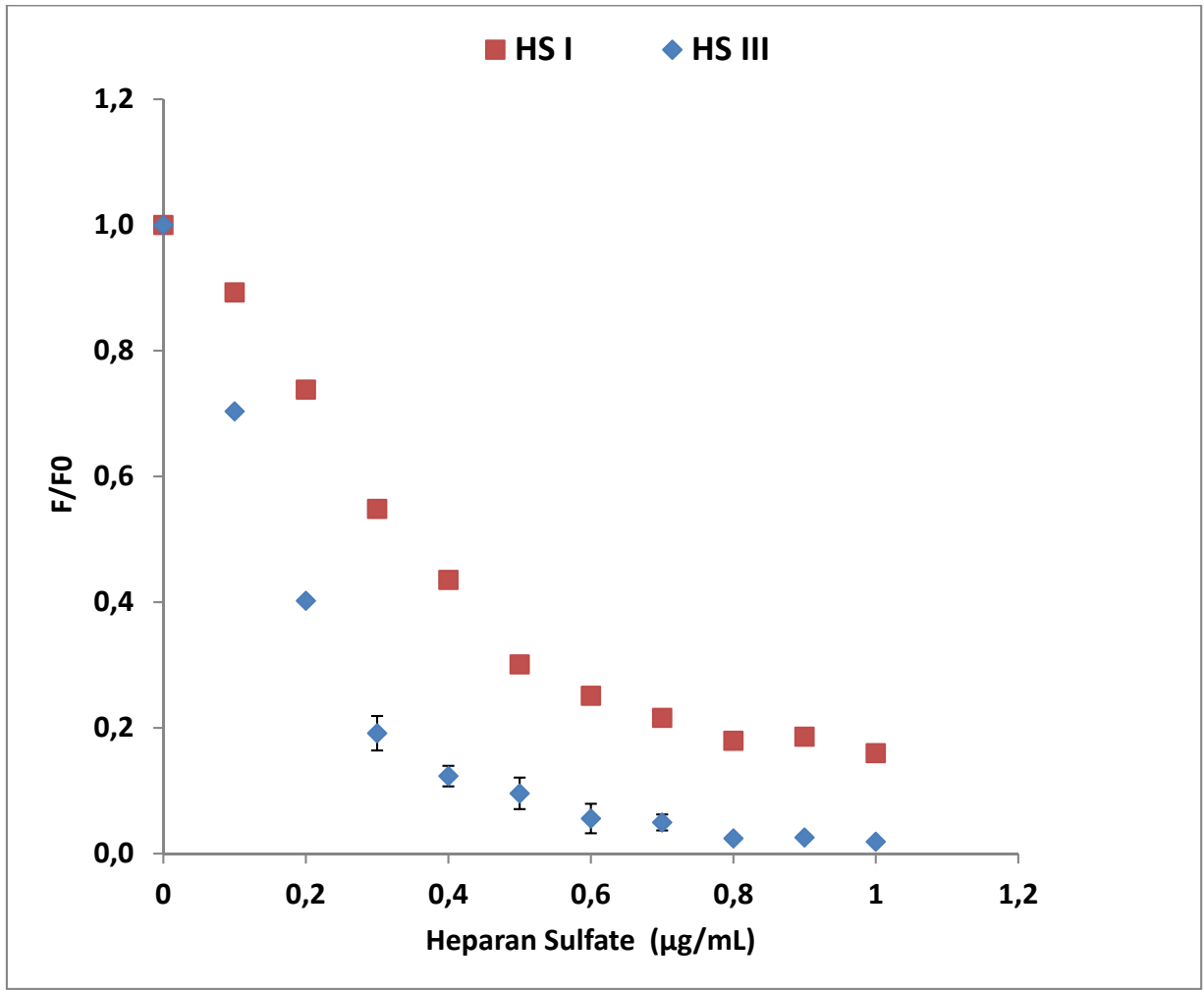

**Figure 2.** Normalised response curve of Heparin Red Ultra using aqueous heparan sulfate samples HS I and HS III in the concentration range 0-1 µg/mL (using an adapted protocol). Manually performed microplate assay. Excitation at 570 nm, fluorescence emission at 605 nm. Averages of duplicate determinations; CV (averaged over all concentrations) for the linear range 0 to ≈0,5 µg/mL of HS I: 0,9 %. CV for the linear range 0 to ≈0,3 µg/mL of HS III: 4,0 %.

*Detection of heparan sulfate in pooled normal plasma*

Pooled normal plasma was spiked with HS I or III and the standard protocol for heparin detection by the Heparin Red Kit applied (figure 3). The stronger response of Heparin Red to HS III than to HS I is confirmed in the plasma matrix. Remarkably, the fluorescence response at low HS levels, i.e. about 5 µg/mL HS III and 2 µg/mL HS I, is comparable to that of the aqueous matrix (figure 1), indicating only little interference by the plasma matrix. Assuming linear response in the range 0-2 µg/mL, the detection limit (3σ of blank, n=6) in pooled normal plasma is 1,1 µg/mL for HS I and 0,5 µg/mL for HS III. At higher HS levels, however, the curve flattens, and an increase from 10 to 15 µg/mL HS triggers only a weak further fluorescence decrease. We interpret this observation by partial but tight association of

Heparin Red with specific plasma components, under formation of a complex in which Heparin Red molecules are not aggregated, i.e. still fluorescent. While the uncomplexed portion of Heparin Red readily aggregates with HS, there might be an equilibrium between the complexed, fluorescent portion of Heparin Red and its nonfluorescent HS aggregate. The stronger the aggregates between Heparin Red and the sulfated polysaccharide are, the better is the linearity of the response also at higher plasma levels, as observed for unfractionated heparin [22] that is expected to form stronger aggregates due to its higher charge density of -1,7 per monosaccharide.

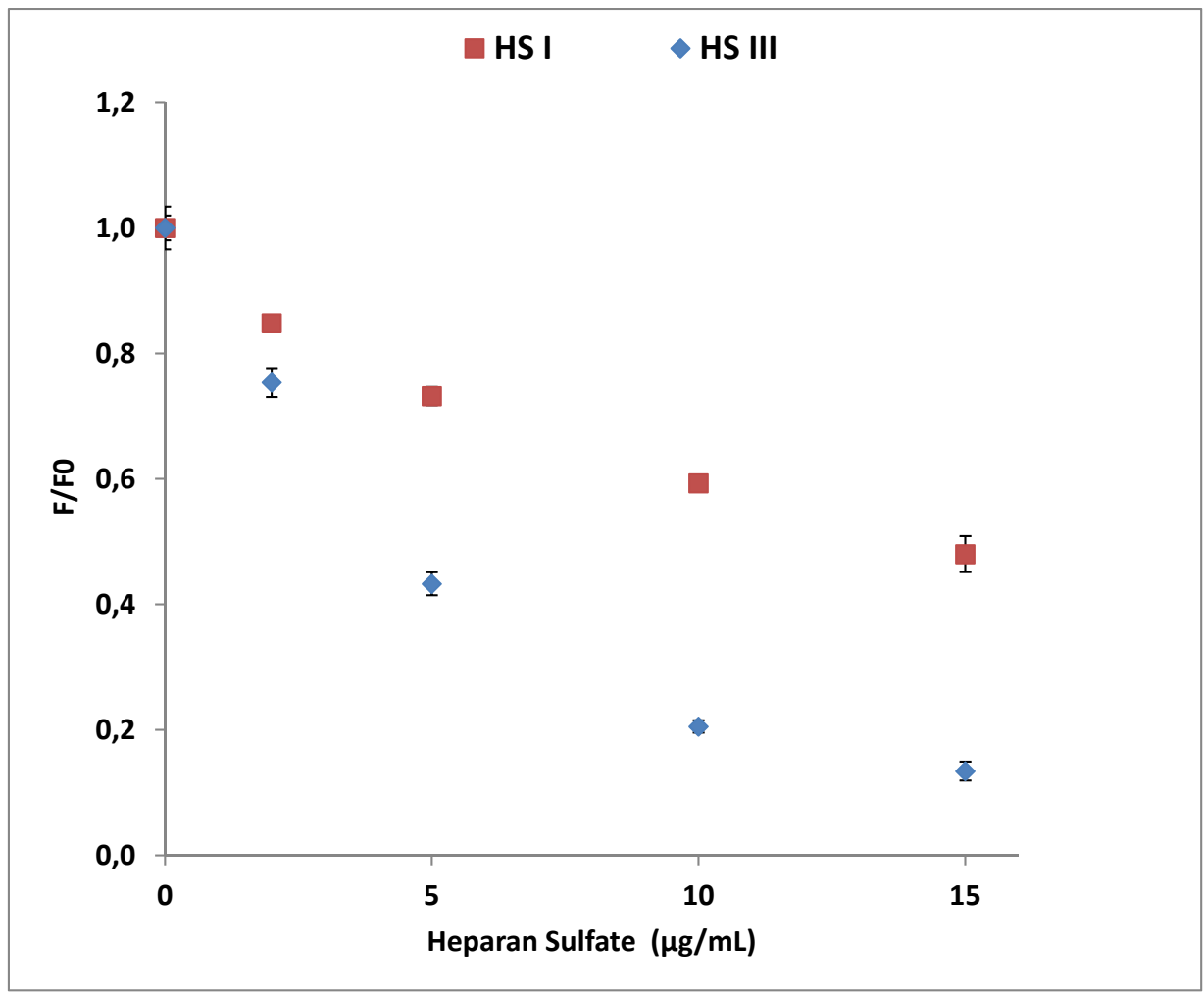

**Figure 3.** Normalized response curve of the Heparin Red Kit using heparan sulfate-spiked pool plasma samples. HS I and HS III in the concentration range 0-15 µg/mL (following protocol of the provider). Manually performed microplate assay. Excitation at 570 nm, fluorescence emission at 605 nm. Averages of duplicate determinations; CV's (averaged over all concentrations) of HS I and HS III: 4,0 %.

*Detection of heparan sulfate in individual normal plasmas*

To explore the influence of plasma matrix variation on precision of the assay, we have spiked plasma samples of 5 healthy donors with HS I and HS III and compared the reponse of the Heparin Red Kit (figure 4 and 5). HS Concetrations in the individual plasmas were calculated by comparison with the pool plasma calibration curves (figure 3) and are listed together with standard deviations in table 3 and 4.

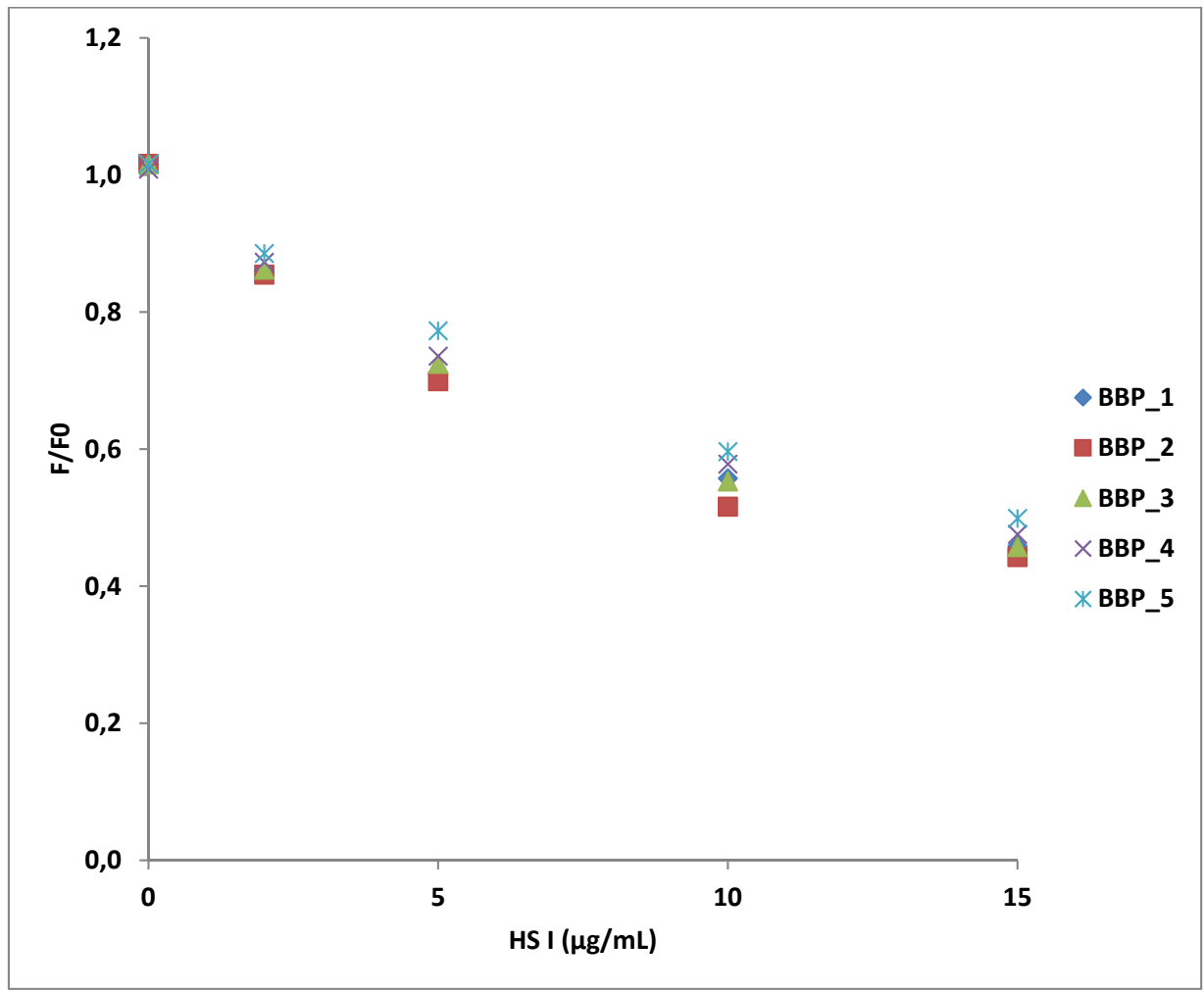

**Figure 4.** Response curves of the Heparin Red Kit using heparan sulfate (HS I)-spiked plasma samples of 5 healthy donors (BBP_1 – BBP_5), following the protocol of the provider. All fluorescence values normalized against the fluorescence of unspiked pool normal plasma (figure 3). Manually performed microplate assay. Excitation at 570 nm, fluorescence emission at 605 nm. Averages of duplicate determinations.

| Spike conc. µg/mL HS I | 0,0 | 2,0 | 5,0 | 10,0 | 15,0 |
|---|---|---|---|---|---|
| Mean (µg/mL) | -0,4 | 2,0 | 5,1 | 10,9 | 17,0 |
| Range (µg/mL) | -0,4 - -0,3 | 1,7-2,3 | 4,0-5,8 | 9,2-13,2 | 14,3-19,3 |
| SD (µg/mL) | 0,04 | 0,20 | 0,64 | 1,33 | 1,70 |

Table 4. Mean, range and standard deviation (SD) for HS I concentrations as determined by the Heparin Red Kit in 5 spiked plasmas of single healthy donors. Concentrations were derived by comparison with the HS I calibration curve for pool plasma (figure 3).

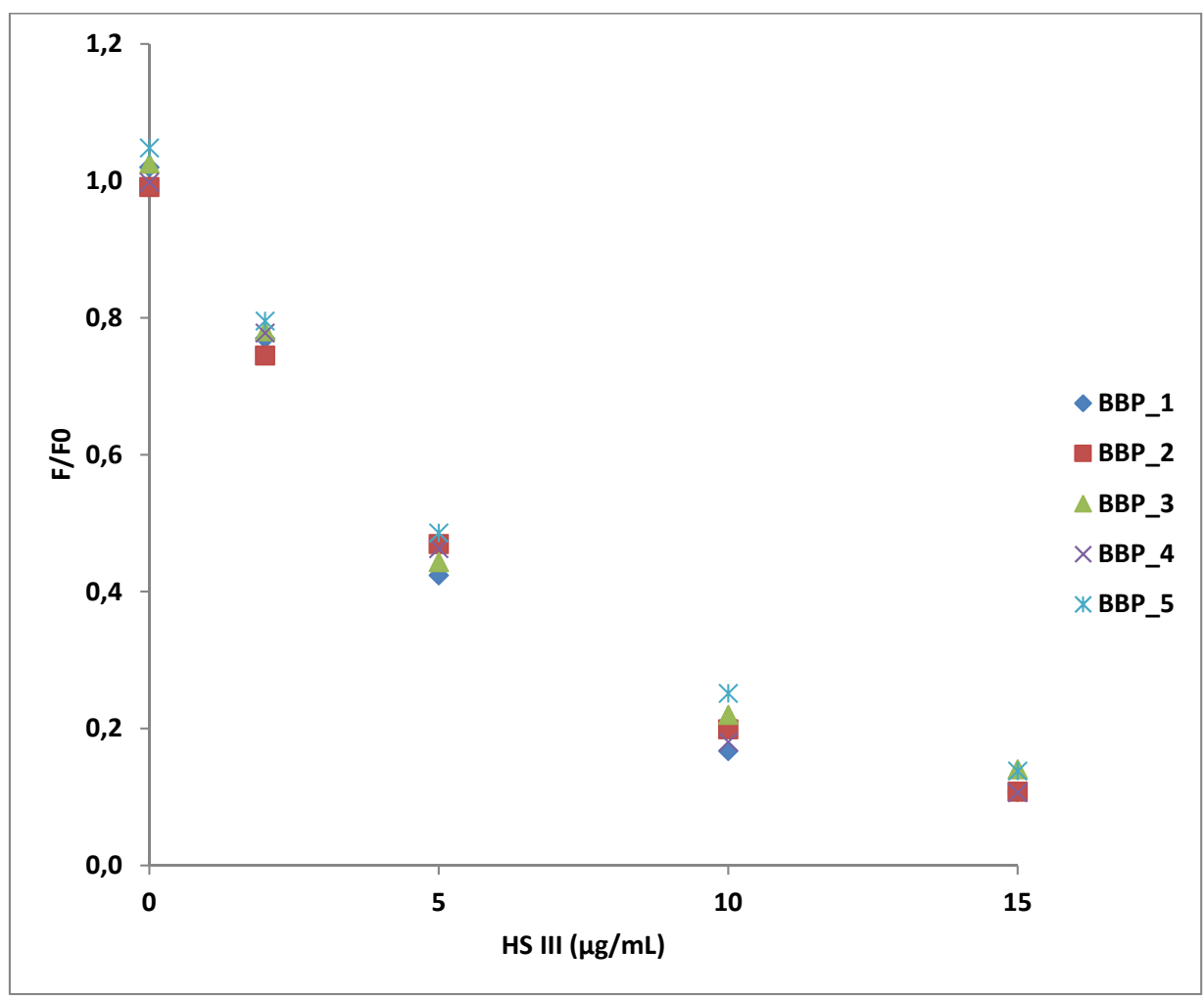

**Figure 5.** Response curves of the Heparin Red Kit using heparan sulfate (HS III)-spiked plasma samples of 5 healthy donors (BBP_1 – BBP_5), following the protocol of the provider. All fluorescence values normalized against the fluorescence of unspiked pool plasma (figure 3). Manually performed microplate assay. Excitation at 570 nm, fluorescence emission at 605 nm. Averages of duplicate determinations.

| Spike conc. µg/mL HS III | 0,0 | 2,0 | 5,0 | 10,0 | 15,0 |
|---|---|---|---|---|---|
| Mean (µg/mL) | 0,0 | 1,6 | 4,9 | 10,6 | 15,4 |
| Range (µg/mL) | -0,2 - 0,1 | 1,4-1,8 | 4,5-5,4 | 8,9-12,1 | 13,7-16,6 |
| SD (µg/mL) | 0,12 | 0,13 | 0,30 | 1,13 | 1,37 |

Table 5. Mean, range and standard deviation (SD) for HS III concentrations as determined by the Heparin Red Kit in 5 spiked plasmas of single healthy donors. Concentrations were derived by comparison with the HS III calibration curve for pool plasma (figure 3).

Interindividual variation is for both HS types significantly higher than the intraassay variability for repeated measurements of the same plasma sample. This indicates a substantial but acceptable influence of the plasma matrix on assay results. Interference by endogeneous plasma glycosaminoglycans is unlikely: In normal plasma, only chondroitin sulfate is present in concentrations that may significantly interfere with the probe. We have already demonstrated [23] by chondroitinase treatment of pooled normal plasma that endogeneous chondroitin sulfates do not affect Heparin Red fluorescence. Partial binding of either the probe Heparin Red (as discussed above) or of HS to specific plasma components may affect probe-target interaction and thus contribute to the interindividual variations of fluorescence response. Heparin Red has a high affinity to heparin and even competes with excess protamine, a strong binder and antagonist of heparin (unpublished results). Nevertheless, certain HS-protein complexes are suggested to have nanomolar dissociation constants and may impede the interaction with Heparin Red [24].

The porcine mucosa heparan sulfate fraction HS I has a charge density of -0,88 per monosaccharide (table 3) and is a good model for human plasma heparan sulfate for which charge densities between -0,55 and -1,1 have been reported (table 1 and 2). Compared with unfractionated heparin [22], a significantly higher mass concentration of HS I is required to trigger 50% fluorescence quenching. This is interpreted with the higher negative charge density of heparin, and the tendency of Heparin Red to form charge neutral aggregates with the polyanionic target [19].

Our experimental data confirm the quantification of HS I in the low µg/mL range by Heparin Red, with a detection limit for pooled normal plasma of about 1 µg/mL. Heparin Red has therefore the potential to detect highly elevated HS levels in the range 1-5 µg/mL that may arise in the course of MPS II, GVHD, Dengue infection or septic shock, by a simple mix-and-read assay.

## Conclusion

Heparan sulfate is a potential biomarker for a variety disease states, but its quantification in complex matrices such as human plasma is challenging. HS levels in plasma are commonly determined by either disaccharide analysis or ELISA; both methods require time-consuming multistep-protocols. The commercially available Heparin Red Kit enables the detection of HS in plasma by a mix-and-read microplate assay within a few minutes. Heparin Red is a fluorescent probe that readily forms a non-fluorescent aggregate with HS and enables direct detection without plasma sample pretreatment. The assay appears to be sufficiently sensitive to detect highly eleveated HS levels in the range 1-5 µg/mL, as reported for mucopolysaccharidosis, graft versus host disease after transplantation, dengue infection or septic shock. The study was performed with spiked plasma samples, and the ability of the assay to detect endogeneous HS in patient samples needs to be confirmed. Our findings mark a significant step toward the development of a convenient and fast method for the quantification of HS in human plasma, with the potential to simplify the detection and advance the acceptance of HS as a biomarker.

**Acknowledgement.** We thank Ms Christina Giese for the preparation of HS plasma spikes.

**Conflict of interest.** R.K. holds shares in Redprobes UG, Münster, Germany. Other authors: No conflict of interest.